\documentclass[aip,pop,reprint,groupedaddress]{revtex4-1}
\usepackage{eurosym}
\usepackage{amssymb,amsmath}
\usepackage{graphicx}

\graphicspath{ {./figures/} }
\begin{document}

\title{Using Poisson-regularized inversion of Bremsstrahlung emission to extract full EEDFs from x-ray pulse-height detector data }
\author{C. Swanson, P. Jandovitz, and S. A. Cohen \\
\textit{Princeton Plasma Physics Laboratory, Princeton University,
Princeton, New Jersey 08543, USA}}
\date{December 7, 2017}
\maketitle

\address{\textit{Princeton Plasma Physics Laboratory, Princeton University,
Princeton, New Jersey 08543, USA}}

\section{Abstract}

We measured  Electron Energy Distribution Functions (EEDFs) from below 200 eV to over 8 keV and spanning five orders-of-magnitude in intensity, produced in a low-power, RF-heated, tandem mirror discharge in the PFRC-II apparatus. The EEDF was obtained from the x-ray energy distribution function (XEDF) using a novel Poisson-regularized spectrum inversion algorithm applied to pulse-height spectra that included both Bremsstrahlung and line emissions. The XEDF was measured using a specially calibrated Amptek Silicon Drift Detector (SDD) pulse-height system with 125 eV FWHM at 5.9 keV.

\section{Introduction}

\label{sec:intro}

X-ray diagnostics used in the study of laboratory and astrophysical plasmas generally fall into two categories: well energy-resolved but narrow energy range spectrometers, such as crystal spectrometers, and poorly energy-resolved but wide energy range spectrometers, such as pulse-height detectors. While crystal spectrometers may achieve relative energy resolution (ratio of resolution to energy) better than $10^{-4}$, they are typically limited to $10^{-2}$ (very narrow) in relative energy range (ratio of lowest energy to highest energy).\cite{bitter} Pulse-height detectors, on the other hand, may achieve $10^3$ in dynamic energy range, capturing useful data from below 100 eV to their maximum value, often well over 100 keV. However, their relative energy resolution is usually limited to $>2\%$ and may be much poorer. \cite{pantazis} 

While diagnostic analyses which rely on atomic line emission of x-rays are well suited by crystal spectrometers, analyses which rely on broad-spectrum emission, such as Bremsstrahlung, are better suited by pulse-height detectors. Furthermore, while line-emission analyses are restricted to a finite set of $N$ degrees-of-freedom, specifically line positions, intensities, and FWHMs, from which to draw $\leq N$ parameters about the target ions, {\it i.e.,} temperatures, densities, and velocities, broad-spectrum analyses have a 1-D continuous degree-of-freedom from which a 1-D continuous parameter such as an energy distribution function may be drawn about the incident electrons. 

One application of pulse-height-detector measurements has been the RHESSI solar observation satellite.\cite{hurford,holman,lin} RHESSI includes nine Germanium pulse-height x-ray detectors (GeD) which are used to energy- and angularly-resolve x-ray emissions from solar flares. To determine the Electron Energy Distribution Function (EEDF) from the X-ray Energy Distribution Function (XEDF), the most accurate and useful algorithm thus far has been the specific variety of Tikhonov-regularized inversion described by Piana, {\it et al.}\cite{piana1994,piana2003} and favorably evaluated by Brown, {\it et al.} \cite{brown} RHESSI represents a best-case scenario for high-quality pulse-height x-ray detectors: a) the RHESSI data is devoid of spectral lines; b) spans three orders-of-magnitude in energy; c) is energy-resolved to better than $1\%$; and d) has enough counts to make counting (Poisson) error a minor part of the total error. 

Our own plasma physics research and other terrestrial fusion-related plasmas provide less ideal conditions for recording pulse-height x-ray spectra. The spectrum may be complicated by spectral lines, lower count rates, and narrower energy ranges than those observed by RHESSI. Under these more challenging conditions and in the range of interest, with a relative energy resolution of $\sim10\%$, there may be several spectral lines obscuring sections of the Bremsstrahlung spectrum, the detector range may span a factor of 100 in energy, and counting statistics may make up the majority of the uncertainty. Because of this, we have developed an algorithm to perform a spectral inversion of XEDF to EEDF which is more resilient to low resolution, spectral lines, and poor counting statistics. The difference in algorithms comes mostly from the cost-function. In Piana's Tikhonov-regularized inversion, a high-dimensional fit is performed to minimize a cost function which corresponds to finding the Maximum \textit{A Posteriori} (MAP) solution given that the measurements are Gaussian random variables. In our Poisson-regularized inversion, a high-dimensional fit is performed to minimize a cost function which corresponds to finding the MAP solution given that the measurements are Poisson random variables with non-negative mean values. Accurate calibration of the detector across the entire energy range is essential.

In this study we use an Amptek X-123 Fast SDD \cite{amptek}, which is a Silicon Drift Detector (SDD) pulse-height x-ray detector to diagnose a plasma with a significant electron component at $\sim 300$ eV. The plasma is produced by a double-saddle antenna at one end of a tandem mirror plasma in the PFRC-II. Up to $\sim 500$ W are transmitted to produce a plasma which has a cold, denser bulk component and a hot, tenuous component. The origin of the hot component is discussed in our previous publication \cite{jandovitz}. Extracting EEDF data from XEDF data is critical to understanding the physics processes responsible for electron confinement, loss, and energization.

\section{Poisson-Regularized Inversion}

\label{sec:inversion}

Poisson-regularized inversion is most easily understood from the direction of Least-Squares method of curve fitting (LSF). It may be shown that the Least Squares solution to a fitting problem is the solution which has the Maximum \textit{A Posteriori} (MAP) probability assuming that the measurements are Gaussian-distributed random variables centered on some function of the model vector.\cite{aster} Bayes's Theorem is used to determine which model vector ($\vec{a}$) has the highest probability given the observation vectors ($\vec{b}$)

\begin{equation}
P(\vec{a}|\vec{b})=\frac{P(\vec{b}|\vec{a})\cdot P(\vec{a})}{P(\vec{b})}
\end{equation}
where $P(q)$ is the probability of event $q$, and $P(q|p)$ is the probability of event $q$ given that event $p$ is true. Interpreting this in the context of spectral inversion, $\vec{a}$ is the vector of discretized EEDF values and $\vec{b}$ is the vector of observed x-ray counts.

This is an optimization problem in $\vec{a}$. Assuming that the prior ($P(\vec{a})$) is uniform, $P(\vec{a}|\vec{b}) \propto P(\vec{b}|\vec{a})$. In LSF, the measurements ($\vec{b}$) are assumed to have Gaussian probability density function centered on some function of the model vector $\vec{l}(\vec{a})$. Interpreting this as a spectral inversion problem, $\vec{l}(\vec{a})$ is the \textit{expected} Bremsstrahlung measurement given a model EEDF, and the measurements are imperfect.

\begin{equation}
\label{eq:gauslikelihood}
P_{Gaus}(\vec{b}|\vec{a}) = \prod_i \frac{1}{\sqrt{2\pi \sigma_i}}e^{-\frac{(b_i-l_i(\vec{a}))^2}{2 \sigma^2_i}}
\end{equation}

As the logarithmic function is monotonic in its argument, maximizing Equation \ref{eq:gauslikelihood} with respect to $\vec{a}$ is the same as minimizing the negative logarithm (the log-likelihood)

\begin{equation}
\label{eq:gauscost}
min(-\ln P_{Gaus}(\vec{b}|\vec{a})) = min(\sum_i \frac{(b_i-l_i(\vec{a}))^2}{2 \sigma^2_i})
\end{equation}

The $\vec{a}$ at which this is achieved is said to be the LSF solution or the $\chi^2$ minimum solution. Tikhonov regularization, as used by Piana for spectral inversion, combines this cost function and a set of priors that allows the absolute amplitude, slope, curvature, etc. to be constrained, if desired.\cite{piana1994, aster}

If we wished to find the equivalent of Equation \ref{eq:gauscost} for measurements $\vec{b}$ distributed as Poisson random variables around some function of the model $\vec{l}(\vec{a})$, we would find that 

\begin{equation}
\label{eq:poislikelihood}
P_{Pois}(\vec{b}|\vec{a}) = \prod_i \frac{[l_i(\vec{a})]^{b_i} e^{-l_i(\vec{a})}}{b_i!}
\end{equation}
\begin{equation}
min(-\ln P_{Pois}(\vec{b}|\vec{a}))=min(\sum_i l_i(\vec{a}) -b_i \ln(l_i(\vec{a})))
\end{equation}
here $\ln b_i!$ has been neglected as it does not contribute to the optimization. 

The XEDF ($\vec{l}(\vec{a})$) is linear in EEDF ($\vec{a}$). Thus, in discretized form the transformation is a matrix $\vec{l}(\vec{a})\rightarrow M\vec{a}$. In the language of matrices, we wish to minimize the following function

\begin{equation}
\label{eq:cost}
C(\vec{a})=\sum_i [(M\vec{a})_i - b_i \ln(M\vec{a})_i] 
\end{equation}
with respect to $\vec{a}$, where $\vec{a}$ is the vector of EEDF values, $\vec{b}$ is the matrix of measured XEDF values, and $M$ is the matrix which transforms EEDF into XEDF.

In practice we solve this optimization problem with a Quasi-Newton method that we implemented in the MATLAB language. As in Tikhonov regularization, add a small ``smoothness'' prior to the cost function, 

\begin{equation}
C_{s}(\vec{a})=\sum_j (\frac{a_{j+1}-a_{j}}{a_j})^2 \frac{E_{corr}^2}{2\Delta E_e^2}
\end{equation}
where $E_{corr}$ is some correlation energy and $\Delta E_e$ is the energy spacing of the EEDF vector. This is a Bayesian prior that neighboring EEDF points will be correlated. We chose $E_{corr}$ to be small, $\sim5$eV, to avoid introducing unphysical results. The addition of this smoothness prior ensures that, in regions of little information about the EEDF, it will behave in a plausible manner. Smoothness is not included in uncertainty computation.

Note that the number of elements in $\vec{a}$, which is the same as the number of columns in $M$, is arbitrary. The error of $a_j$ values will be discussed shortly. Choosing a small value of the dimensionality of $\vec{a}$ will decrease the uncertainty of each $a_j$ but not resolve features of the derived EEDF. Choosing a large value of the dimensionality of $\vec{a}$ will increase the uncertainty of each $a_j$ but resolve better the energy-space of the EEDF. Natural choices for the dimensionality of $\vec{a}$ are the number of measured points in $\vec{b}$ for maximum resolution, or the number that gives the energy spacing of $\vec{a}$ the same resolution as the energy-resolution of the measurement, to balance resolution and uncertainty.

This Poisson-regularized inversion is well-suited to XEDFs that suffer from poor counting statistics. If Tikhonov-regularized inversion is used in cases of low signal-to-noise ratio (SNR), positive-definiteness is not assured and the derived EEDF can be negative, a non-physical result. This is because the assumption that the measurement is Gaussianly distributed is only justified at high numbers of x-rays counted. 

Poisson-regularized inversion is also well-suited to low-energy-resolution data. In pulse-height x-ray detectors, resolution is limited by the counting statistics of the number of electron-hole pairs which are produced by the incident x-ray.\cite{pantazis} Thus, the response of the detector to a monoenergetic beam of x-rays is a mostly-Gaussian distribution of energies centered on the correct energy. This response function can be made into a response matrix when the response functions to many different energies are concatenated as the columns of a matrix. This will be discussed later. Multiplying this matrix $M_{res}$ into the EEDF - XEDF transformation $M_{Brem}$ yields a resolution-included response matrix $M_{res\cdot Brem}$ which allows the resolution of the detector to contribute to the derived EEDF and its reported uncertainties in a correct and self-consistent way. This procedure is not practiced in the most widely used Tikhonov-regularized inversions.\cite{piana2003}

Poisson-regularized inversion also yields natural uncertainties and error bars in a self-consistent way that Tikhonov-regularized inversion does not. The systematic uncertainty of each point $a_j$ may be determined by finding the value of $a_j+\Delta a_j$ which increases $C(\vec{a})$ by the value $1/2$, equivalent to a $1\sigma$ deviation for Gaussian statistics. The statistical uncertainty of a section of the spectrum may be found the same way, except all points except $a_j$ are then optimized to minimize $C(\vec{a})$. It is this new $C$ which is must be larger than the old $C$ by $1/2$.

\section{Detector Response Function}

\label{sec:response}

The EEDF derived from Equation \ref{eq:cost} is sensitively dependent on the transformation matrix $M$. We call this matrix the response matrix, as it is the detector's response to the presence of a mono-energetic EEDF.

$M$ consists of three response matrices, multiplied together. 

\begin{equation}
M=M_{res} \cdot M_{trans} \cdot M_{Brem}
\label{eq:M}
\end{equation}
where $M_{Brem}$ is the Bremsstrahlung production matrix, $M_{trans}$ is the window transmission matrix, and $M_{res}$ is the resolution response matrix. They must be multiplied in this order, as these transformations are not commutative.

If the physics and detector are well known, then $M$ is specifiable analytically. In practice, all or some of $M_i$ matrix factors in Equation \ref{eq:M} must be measured.

\subsection{$M_{Brem}$}

$M_{Brem}$, a discretized transformation function, multiplies the EEDF to determine the XEDF that is produced by the plasma. This is relatively well-known in the non-relativistic case in which the Born approximation holds.\cite{elwert, hutchinson} The power into $\nu \rightarrow \nu+d\nu$ emitted by one electron interacting with a field of Coulomb potentials is

\begin{equation}
\label{eq:bremtransfreq}
\frac{dP}{d\nu}=\frac{32\pi^2}{3\sqrt{3}}n_i Z^2 r_e^3 m_e c^2 \frac{c}{v_1} G(v_1,v_2)
\end{equation}
where $r_e$ is the classical electron radius, $m_e$ is the electron mass, $n_i$ is the number density of Coulombic scattering centers, $Z$ is the charge of the scattering center, $\nu$ is the frequency of the x-ray, and $G$ is the Gaunt factor, which in the energy range of interest can be approximated \cite{elwert} 

\begin{equation}
\label{eq:elwert}
G=\frac{\sqrt{3}}{\pi}\frac{\nu_2[1-e^{-2\pi \nu_1}]}{\nu_1[1-e^{-2\pi \nu_2}]} \ln\frac{\nu_2+\nu_1}{\nu_2-\nu_1}
\end{equation}
where $\nu_i^{-2}=\frac{\frac{1}{2}m_e v_i^2}{Z^2R_y}$, $v_1$ and $v_2$ are the electron velocities before and after the emission, and $R_y$ is Rydberg's constant. This is true only of fully ionized targets. Partially ionized targets will be considered in Section \ref{sec:partialion}.

We are interested in the number of x-rays within an energy range $E_x \rightarrow E_x+\Delta E_x$ that hit an active area within a time period $\tau$ from a plasma with electron density $n_e$ and EEDF $f(E_e)$. Transforming Equation \ref{eq:bremtransfreq} into these quantities, we get

\begin{multline}
\label{eq:bremtrans}
b_i=\tau V \Delta E_x n_i n_e Z^2 \frac{32\pi^2}{3\sqrt{3}} \frac{(m_e c^2)^{3/2}}{h} r_e^3 \times \\ \int dE_e \frac{f(E_e) G(E_e,E_{x,i})}{E_{x,i}\sqrt{E_e}}
\end{multline}
where $b_i$ is the number of x-rays that enter the detector in energy bin $i$, $\tau$ is the time that data is collected, $V$ is the effective volume from which every x-ray is collected, $\Delta E_x$ is the interval of energy that each energy bin spans, $E_{x,i}$ is the central energy of x-ray energy bin $i$. 

Posing Equation \ref{eq:bremtrans} in the language of transformation matrices, 

\begin{equation}
\label{eq:M_Brem}
M_{Brem,i,j}=K \Delta E_e \frac{a_j G(E_{e,j},E_{x,i})}{E_{x,i} \sqrt{E_{e,j}}}
\end{equation}
\begin{equation*}
\vec{b}_{Brem}=M_{Brem}\vec{a}
\end{equation*}
where $K$ is the constant before the integral in Equation \ref{eq:bremtrans}, $\Delta E_e$ is the discretization interval for $E_e$, $a_j$ is the $j$th element of the discretized EEDF vector, and $E_{e,j}$ is the electron energy associated with $a_j$. 

\begin{figure}[tbp]

\centering\includegraphics[scale=0.45]{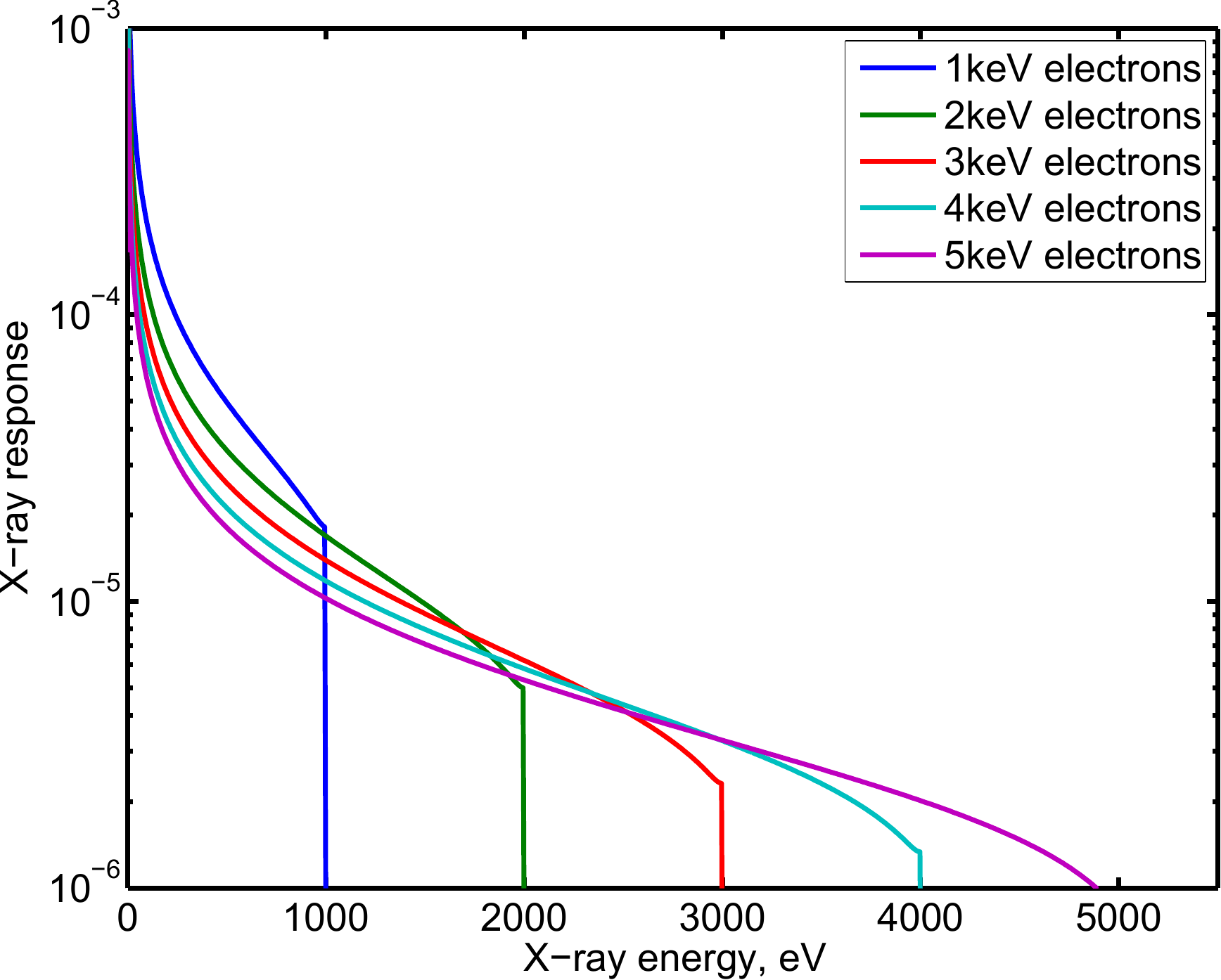}
\caption{Example $\vec{b}_{Brem}$ vectors produced by monoenergetic EEDFs, as predicted by Equation \ref{eq:M_Brem}. $M_{Brem}$ is zero where $E_{x,i}>E_{e,j}$, as an electron cannot produce an x-ray with more energy than itself.}
\label{fig:elwert}
\end{figure}

Examples of the Bremsstrahlung response from monoenergetic electrons as predicted by Equation \ref{eq:M_Brem} are depicted in Figure \ref{fig:elwert}.

\subsection{$M_{trans}$}

$M_{trans}$ is the transmission efficiency matrix that transforms the XEDF produced by the plasma to the XEDF which penetrates the detector window and is incident on the detector. As this process does not change the energy of the x-rays, $M_{trans}$ is a diagonal matrix. The value of element $M_{trans,i,i}=\vec{m}_{trans,i}$ is the value of the transmission efficiency of the window and detector at energy $E_{x,j}$.

This transmission efficiency is published for each pair of Amptek detectors and windows by Amptek. \cite{amptek} We used a X-123 FAST SDD detector with a Si$_3$N$_4$ C1 window. The Amptek C1 window is composed of a 90-nm layer of silicon nitride, a 250-nm layer of grounded aluminum, and a $78\%$-open grid made of 15-$\mu m$-thick silicon.  The SDD is 500-$\mu m$ thick which limits the detector's upper energy. 

\begin{figure}[tbp]
\centering\includegraphics[scale=0.5,trim={0 0 6cm 0},clip]{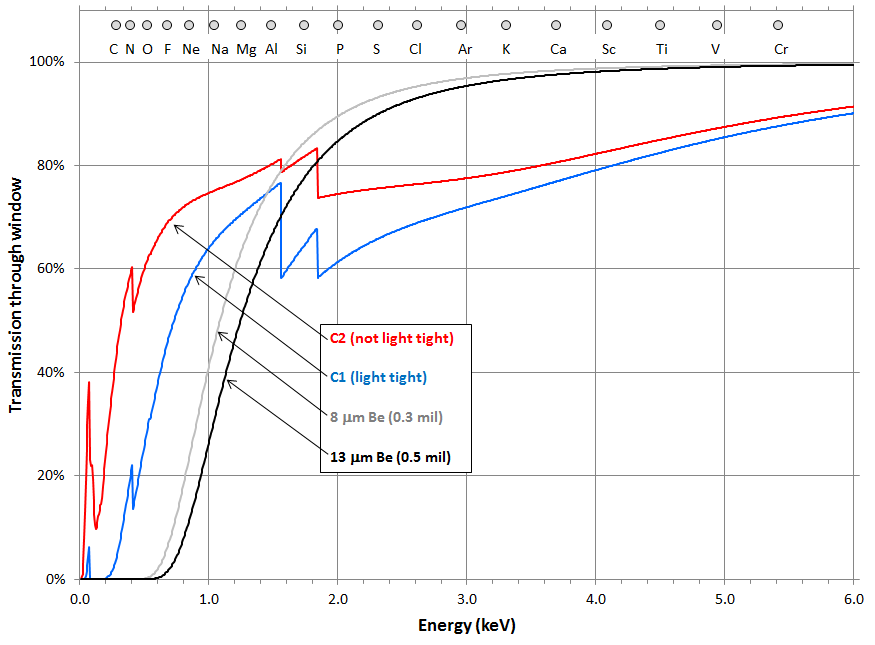}
\caption{Published transmission efficiency of an Amptek SDD detector with various windows as a function of incident x-ray energy. \cite{amptek} We used a C1 window.}
\label{fig:trans_amptek}
\end{figure}

The transmission efficiencies of this combination of window and detector are published, as depicted in Figure \ref{fig:trans_amptek}, or can be calculated from the given dimensions using NIST's database of x-ray attenuation coefficients, which tabulate calculations of Seltzer.\cite{NIST, seltzer}

\begin{equation}
M_{trans}=Diag[\vec{m}_{trans}]
\end{equation}
where $\vec{m}_{trans}$ is the vector of transmission efficiencies at energies $\vec{E_x}$. 

The windows are manufactured with a $\pm10\%$ thickness tolerance of the silicon nitride film; when instructing how to determine line amplitudes for x-ray material assaying, Amptek instructs the user to first calibrate with a known source. In Section \ref{sec:calibration} we extend this procedure to broad-spectrum x-ray emissions.

\subsection{$M_{res}$}

The resolution of pulse-height detectors are limited by two phenomena: electrical (thermal, shot, etc.) noise of the counting electronics and counting statistics of the finite number of electron-hole pairs produced by the incident x-ray. 

This latter produces a full-width-at-half-maximum (FWHM) which is proportional to the square root of the signal. This is so-called Fano noise \cite{fano} and, in the Amptek X-123 FAST SDD detector, has form\cite{amptek}

\begin{equation}
FWHM_{Fano}=\sqrt{2.404eV \cdot E}
\end{equation}
where all quantities are in eV.

The electrical noise is constant with respect to incident x-ray energy. According to the Amptek calibration procedures, the response function to a monoenergetic x-ray distribution is mostly Gaussian centered at the x-ray energy and with FWHM contributions from electrical and Fano noise.\cite{amptek} The FWHM of the electrical noise is measured each time data is recorded by first recording the measured XEDF from no x-ray signal, yielding a Gaussian near 0 eV with the correct FWHM. 

\begin{equation}
FWHM(E_x)=\sqrt{FWHM^2_{elec}+FWHM^2_{Fano}(E_x)}
\end{equation}

Thus the resolution response matrix $M_{res}$ which transforms the XEDF vector which is incident on the detector into the measured XEDF vector is many column vectors of discretized Gaussian values concatenated into rows, the central values and FWHMs of which depend on the row.

\begin{equation}
M_{res,i,j}=\frac{1}{\sqrt{2\pi\sigma(E_{x,j})}}e^{-\frac{(E_{x,i}-E_{x,j})^2}{2\sigma^2(E_{x,i})}}
\end{equation}
where $\sigma(E_{x,i})=FWHM(E_{x,i}) /(2\sqrt{2\ln 2})$.

\section{Comparison with Piana \textit{et. al.}}
\label{sec:comparison}

\begin{figure}[tbp]
\centering\includegraphics[scale=0.45]{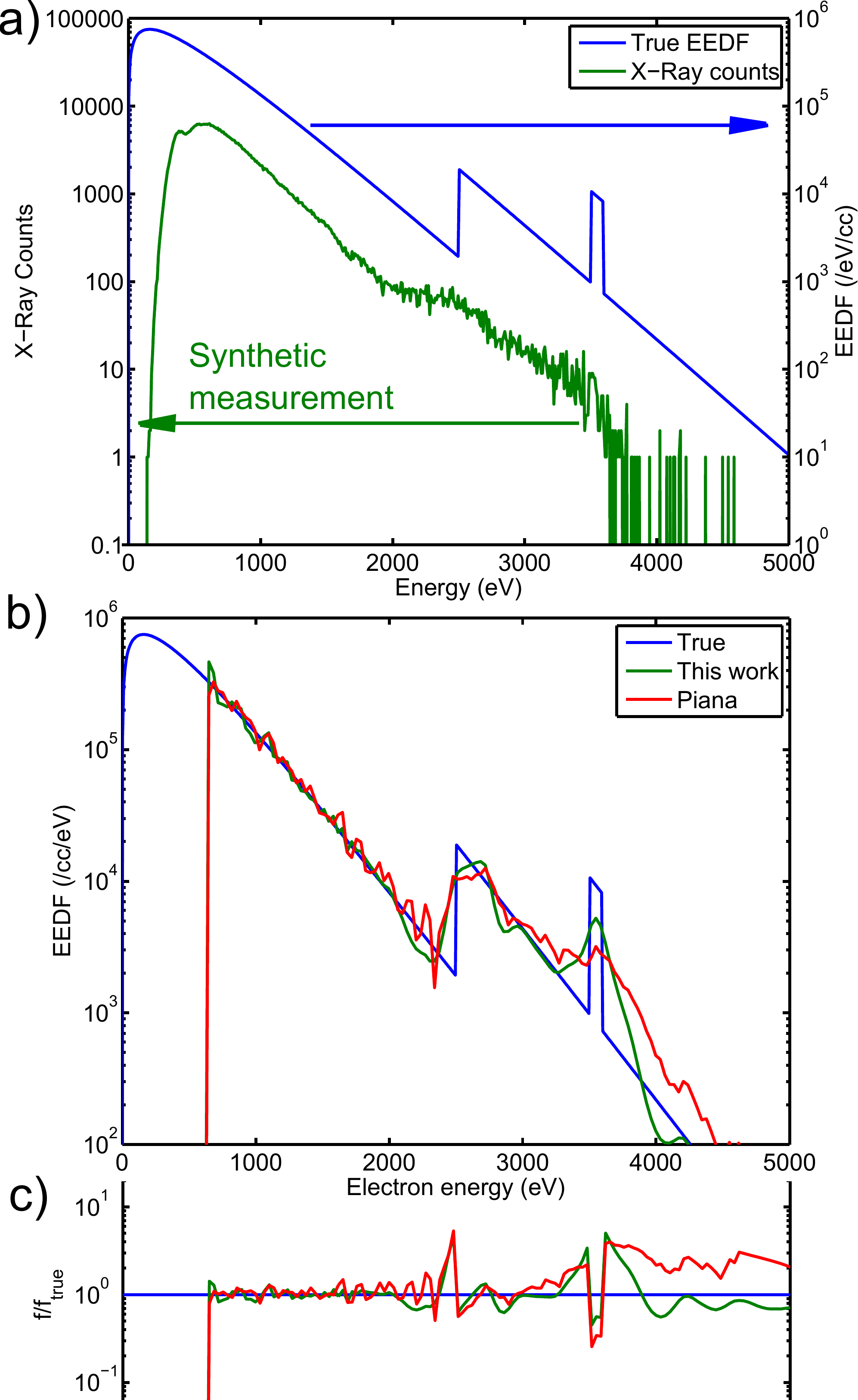}
\caption{An example comparison of Poisson-regularized Inversion and the Piana algorithms for recovering an EEDF. a) The true, supposed EEDF and the synthetic x-ray data generated from it. b) Computed EEDFs compared to the true EEDF. c) Discrepancy factor between each computed EEDF and the true value.}
\label{fig:comparison}
\end{figure}

In order to illustrate the differences between the Poisson-regularized inversion of this paper and the Tikhonov-regularized inversion of Piana \textit{et. al.}, we supposed a functional form for an EEDF, depicted in Figure \ref{fig:comparison}a. From this EEDF, we generated a synthetic XEDF by applying the transformation matrix we found in Section \ref{sec:response}, $\vec{m}_{true}=M\vec{f}_{true}$. For each energy bin $E_{x,j}$, a random value was generated from a Poisson distribution centered at $m_{true,j}$. This synthetic XEDF, $m_{Poisson}$, should be thought of data that could plausibly have come from a laboratory plasma. It incorporates the effect of Bremsstrahlung, window transmission, finite resolution, and counting statistics. 

The specific functional form of $f_{true}$ was based on a Maxwellian distribution with density $n_e=5\times 10^8 /cc$ with temperature 320eV. We assumed a target of 0.30 mTorr Hydrogen gas. Data was ``accumulated'' for 45 minutes. These are all plausible conditions for the PFRC-II experiment. As can be seen in Figure \ref{fig:comparison}, both the Poisson-regularized inversion and that of Piana captured this behavior as far as it persists in energy. Piana's EEDF exceeds $30\%$ discrepancy at 1600eV. Poisson-regularized EEDF does not exceed $30\%$ discrepancy until 2500eV, where the Maxwellian character ends.

To demonstrate resolution of sharp features, we added a jump in EEDF at 2500eV. As can be seen in Figure \ref{fig:comparison}a, counting statistics (Poisson error) have become significant in this energy range. Despite this complication, both computed EEDFs show a jump of comparable amplitude at 2500eV, with similar artifacts due to resolution. 

To demonstrate sensitivity, a tenuous beam was added. We chose 3550eV with density $n_{beam}=9\times 10^5 /cc$. In this energy range, the Poisson error in the synthetic XEDF are extreme, with variance $> 35\%$. There were only 67 synthetic x-rays ``measured'' above 3550eV. The Poisson-regularized EEDF clearly shows a beam centered at 3550eV, however the density excess is $15\%$ higher than the true density of the beam, and the FWHM of the beam is 250eV, larger than the 100eV of the true beam. However, the EEDF calculated from the algorithm of Piana showed no detectable beam-vs-background relationship from which a density could be calculated.

The ability to detect a beam of $\sim 0.1\%$ of the bulk density is potentially of great use. A beam of these beam parameters can be injected from an electron gun, providing a valuable diagnostic of confinement time and slowing down profile. A beam of these parameters can be a source of free energy for instabilities such as two-stream instability.

 These data confirm the claim in Section \ref{sec:inversion} that Poisson-regularized inversion is more suitable than Tikhonov-derived inversions in the case that the relevant EEDF feature is in an energy range in which uncertainties and resolutions are high. When the signal-to-noise ratio is low, the Tikhonov-derived Piana inversion produces unreliable results where Poisson-regularized inversion reproduces the energies and amplitudes of important EEDF features.

\section{Calibration}
\label{sec:calibration}

\begin{figure}[tbp]
\centering\includegraphics[scale=0.58]{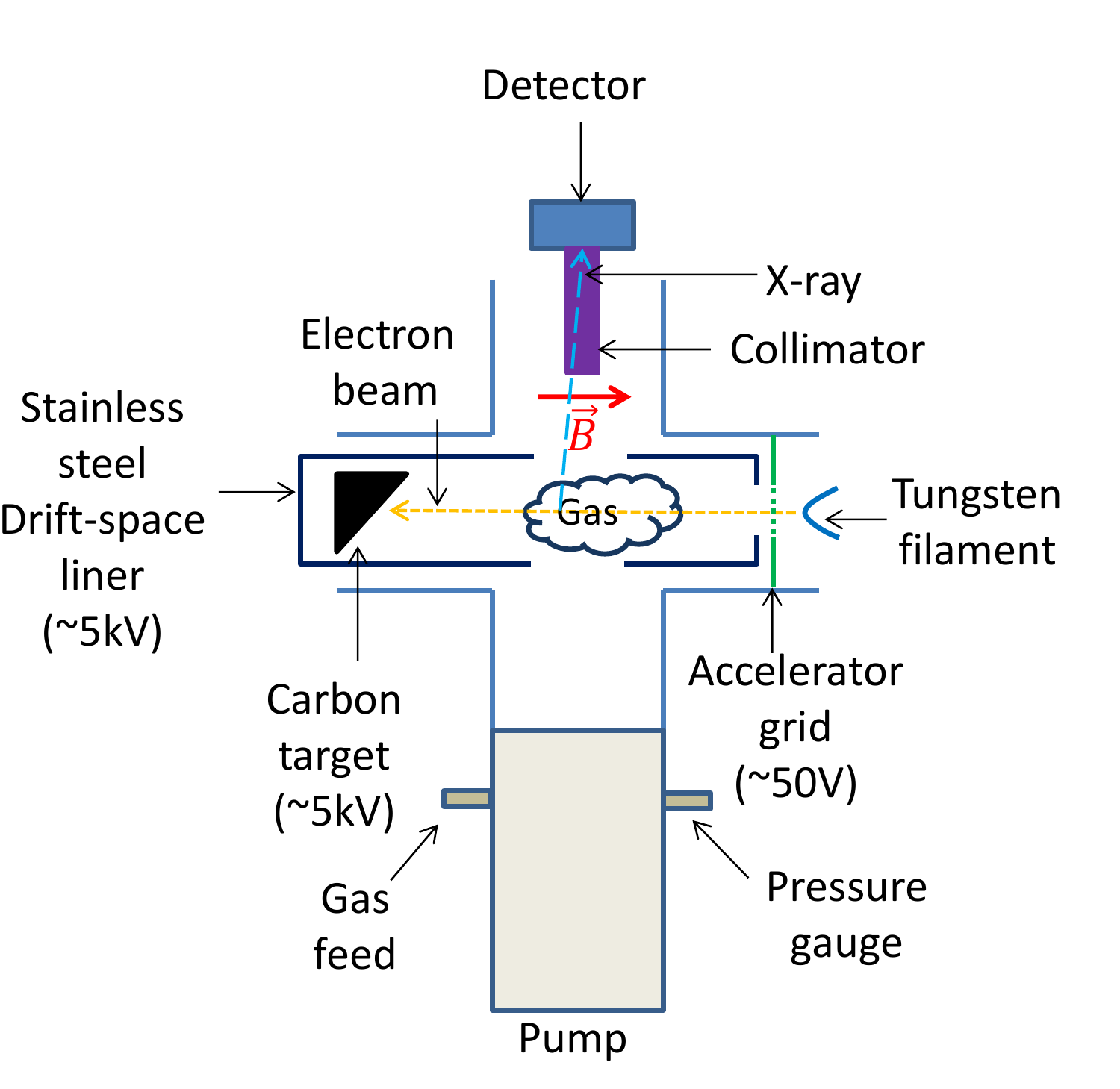}
\caption{The gas-target x-ray tube used for calibration of the detector.}
\label{fig:chaostube}
\end{figure}

To calibrate the SDD, a gas-target x-ray tube was built using a 20-cm total length Pyrex cross with 5-cm inner diameter, depicted in Figure \ref{fig:chaostube}. At the far right port of the cross, a 0.010'' diameter tungsten filament was heated by a DC power supply to thermionically emit electrons. The potential along the filament's length varied about 3 V with respect to ground. The electrons were extracted by a nearby stainless steel acceleration grid at +50 V and then accelerated into the gas target volume. At the left port of the cross, a triangular prism carbon target was biased up to 5 kV. So that the electrical potential and electron energy were constant in the gas target volume, a stainless steel drift-space liner was placed around the gas-target volume, with two 5-cm holes at the top and bottom to allow line-of-sight of the detector through the liner and pumping. This drift-space liner was biased at the carbon target voltage. Wrapped around the horizontal arms of the cross were 25 turns each of wire carrying 4 A of current, providing a $\sim4G$ axial magnetic field to contain the electron beam. The current collected by the carbon target and drift space liner was typically $100\mu A$.

At the bottom port of the cross, gas was fed into the gas target volume by an adjustable needle valve. A Leybold-Heraeus Turobovac 150 turbomolecular pump maintained a base pressure below $10^{-5}$ Torr, as measured by a Granville-Phillips ion gauge. The gas pressure was kept low enough that the thermionic current was not noticeably supplemented by current from ionization of neutral gas. 

At the top port of the cross, the X-123 FAST SDD detector was mounted, with a grounded stainless steel 3-baffled collimator to restrict line-of-sight.

\subsection{Energy calibration}

The energy scale of each detector must be calibrated for each combination of pulse-shaping parameters. At its most basic level, the data obtained in every run is a list of counts per energy channel, $\vec{b}$. The energies to which they correspond and the width of the energy bin are not specified. Amptek instructs the user to use known, calibrated signals from radioisotopes such as Iron-55 to calibrate this energy scale.\cite{amptek}

We supplemented this by using the gas-target x-ray tube that we constructed for this purpose. Using different fill gases with different $K-\alpha$ line energies we were able to calibrate the energy scale over a large range of energies, from the $K-\alpha$ line of Carbon at 277eV to the $K-\alpha$ line of Argon at 2958eV. We considered the position of six elemental $K-\alpha$ lines: Carbon from methane gas fill, Nitrogen from air gas fill, Oxygen from air gas fill, Neon from neon gas fill, Aluminum from x-ray fluorescence from aluminum foil, and Argon from air gas fill.

 The detector has configurable time constants, including pulse rise time and flat top, that affect the tradeoff between resolution and count rate \cite{amptek}. The pulse peaking time and flat-top time also affected the energy calibration of the detector. We calibrated the energy scale of the detector under 7 peaking and flat-top time combinations applicable to our experiment. In each case the calibration function was found to be linear; that is the $N$th energy bin corresponded to energy $E=A+BN$. Energy calibrations performed months apart varied by $<0.5\%$.

\subsection{Transmission calibration}

\begin{figure}[tbp]
\centering\includegraphics[scale=0.45]{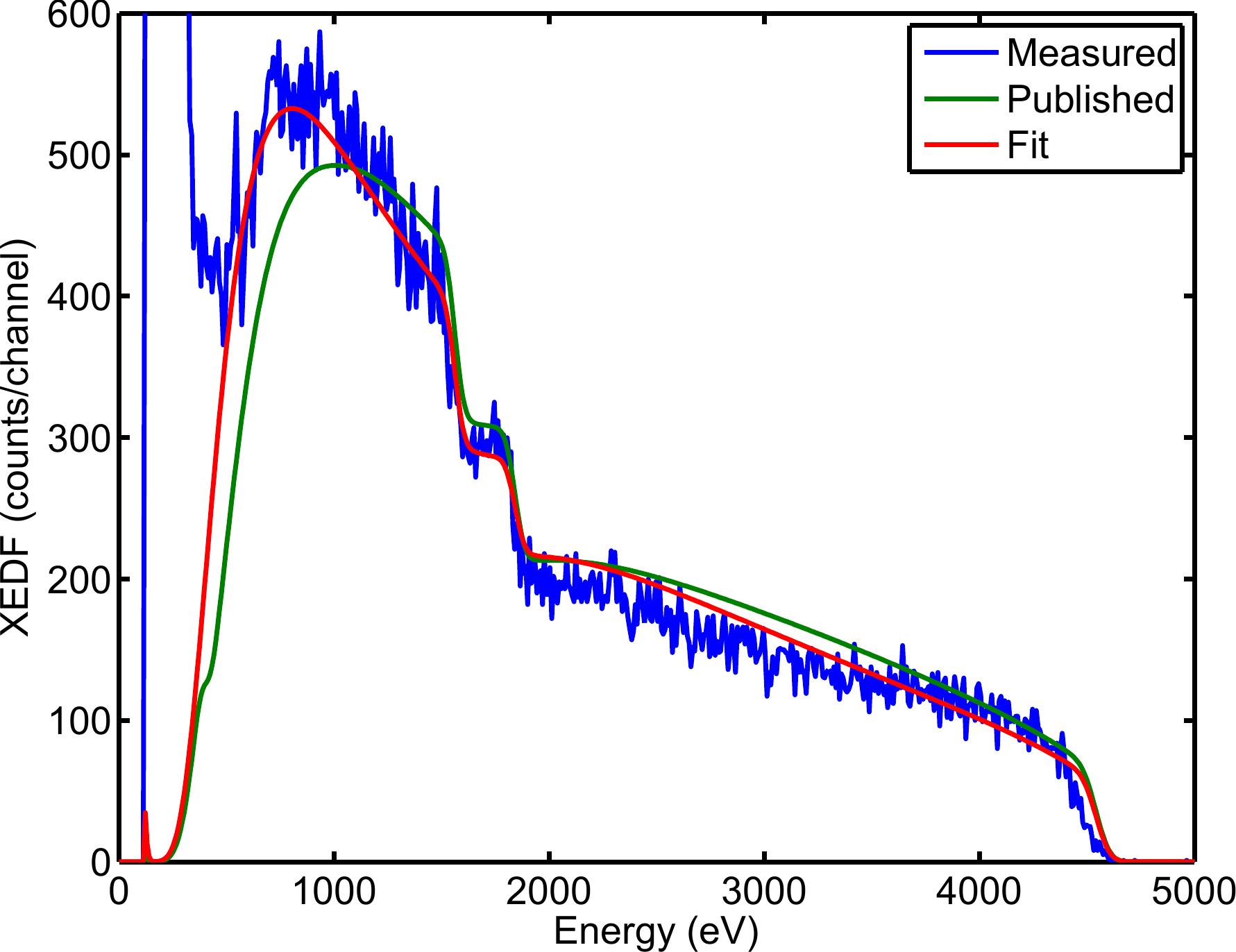}
\caption{Transmission Calibration. In blue, the XEDF as directly reported by the SDD. In green, the XEDF expected from a monoenergetic EEDF assuming the published transmission efficiencies. In red, assuming the transmission efficiency produced by the fit parameters given in the text.}
\label{fig:transcal}
\end{figure}

As depicted in Figure \ref{fig:transcal}, we recorded the XEDF from the monoenergetic EEDF produced by the gas-target x-ray tube. Its shape differed slightly from the shape expected from the published x-ray transmission efficiency depicted in Figure \ref{fig:trans_amptek}. This is because the C1 window used in our detector was manufactured to some finite tolerance, which each of its thicknesses subject to some variation. The four free parameters that characterize the C1 window are: Silicon Nitride layer thickness (nominally 90nm), Aluminum backing thickness (nominally 250nm), Silicon grid thickness (nominally 15$\mu$m), and Silicon grid open fraction (nominally $78\%$).

As depicted in Figure \ref{fig:transcal}, thicknesses and open fractions slightly different than this give improved fit with the measured XEDF. The best fit was produced by Silicon Nitride layer of 64nm, Aluminum layer of 244nm, Silicon grid thickness of 23$\mu$m, and Silicon grid open fraction of $80\%$.

The agreement between the calculated monoenergetic XEDF and measured XEDF gives us confidence in our Bremsstrahlung production model.

\section{Spectral Lines}
\label{sec:spectral}

\begin{figure*}[tbp]
\centering\includegraphics[scale=0.5]{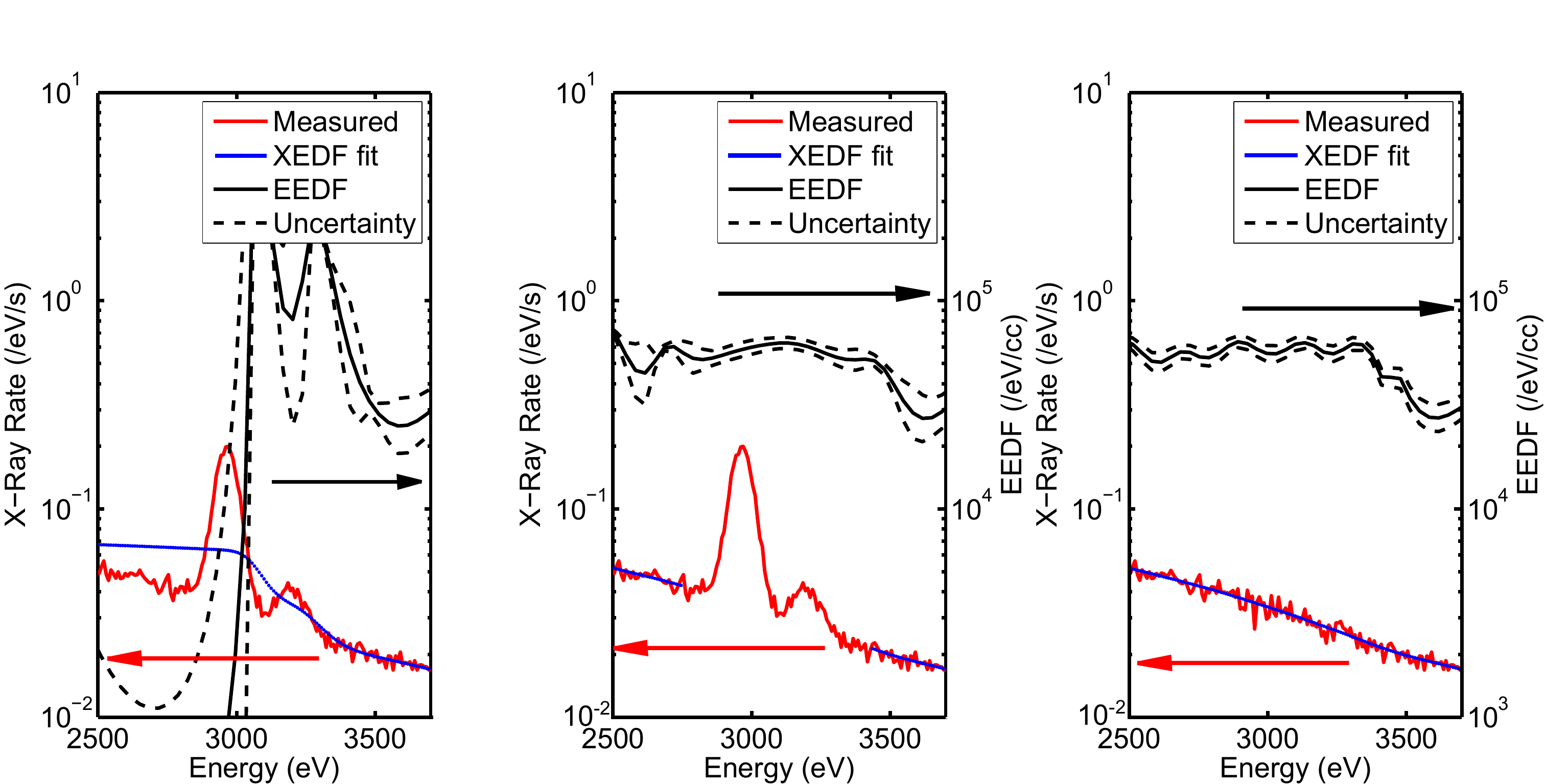}
\caption{Illustration of two Argon spectral lines, their effect on derived EEDF, and two potential methods of correcting for them. a) shows the effect of an un-corrected spectral line. This spectrum cannot be the result of Bremsstrahlung processes, so the fit fails and the resultant EEDF has large non-physical features. b) shows the method of correction used in this paper: the elements of $i$, the discretized energy vector, which correspond to the energy range obscured by the spectral line have been removed from the sum in Equation \ref{eq:cost}. c) shows the method of manually subtracting a Gaussian function from the observed x-ray spectrum until the result appears subjectively correct. }
\label{fig:wrong-spectral-lines}
\end{figure*}

$M_{Brem}$ from Equation \ref{eq:M_Brem} was formulated only considering x-rays from Bremsstrahlung emission. Because of this, the presence of line radiation in the energy domain-of-interest will cause the algorithm to fail. See Figure \ref{fig:wrong-spectral-lines}a.

Because of the large, obvious unphysical features in the un-corrected EEDF, this method can be used to find spectral lines. The failure of a fit indicates that the measured x-ray spectrum could not have been produced by Bremsstrahlung only, and that a spectral line must be present.

To invert XEDFs with prominent spectral lines in the domain-of-interest, we also used the following procedure: We determined the energy range over which the spectral line extended. This corresponds to some range of $i$ values in Equation \ref{eq:cost}. We removed this interval from the summation in that equation. This method removes the unphysical features of the spectral line, and produces a plausible but uncertain EEDF in the vicinity of the excised line. To understand why this is the case, we must first discuss some of the properties of the reconstructed EEDF's dependence on the measured XEDF.

Normally the EEDF in some energy range $E_i$ is determined mostly by the local behavior of the XEDF in the vicinity of $E_i$ and partly by the behavior of the XEDF at lower energy than $E_i$. Both behaviors of the XEDF are ``evidence'' in the Bayesian sense of the word which specify the value and uncertainty of the EEDF. If a spectral line is excised with this method, however, EEDF values are determined by the XEDF only at lower energies. These EEDF points have larger uncertainty because the lower-energy XEDF points are less sensitively dependent on these EEDF values. Large EEDF features may produce only small XEDF changes when the energy range of interest is excised. This method is depicted in Figure \ref{fig:wrong-spectral-lines}b.

Spectral lines can also be removed ``manually," by choosing for each line a candidate peak location ($E_p$), height ($h_p$), and width ($w_p$), subtracting that peak from the measured XEDF and varying $E_p$, $h_p$, and $w_p$ to minimize the XEDF deviation from a ``smooth line" in the vicinity of the peak. Essentially, the resultant XEDF should appear, ``by eye,'' as though there were no spectral line. Because of the shape of $M$, small differences in the XEDF correspond to large differences in the EEDF, so care must be taken not to introduce artificial excursions from the correct EEDF. This method has the benefits of corroborating the SDD's energy resolution and measuring the impurity content. This method has the drawback that human bias is introduced. This method is depicted in Figure \ref{fig:wrong-spectral-lines}c.

The Gaussian peaks removed to create Figure \ref{fig:wrong-spectral-lines}c were centered at 2964.5 eV and 3196.6 eV, very close to Argon spectral lines, indicating that our energy calibration is correct to $0.2\%$. The peaks had FWHMs of 91.4 eV and 93.4 eV, within Amptek's specifications for this range. The XEDF falls over this energy range while the EEDF appears flat. This is not a contradiction, as a constant EEDF means that more and more electrons can produce x-rays of a given energy as that energy decreases. The dip in EEDF at 3600 eV may be due to an effect discussed in Section \ref{sec:partialion}: the complicated behavior of the Gaunt factor of Avdonina and Pratt around an electron shell energy\cite{avdonina}. We will explore this Argon-impurity discharge more in a later publication.

The most correct way to add spectral line considerations is to include a model for their production from a general EEDF. This model would modify the x-ray response function $M_{Brem}$. This requires the concentrations of the emitting gases to be known. For our plasmas, the correct model is the collisional radiative model, which is non-linear in the EEDF and requires knowledge of particle confinement, cannot be subjected to this analysis. We did not implement this feature.

\section{Partially ionized high-Z targets}
\label{sec:partialion}

For all data and analyses presented in this paper, we do not consider the presence of other electrons bound to the target species. When we consider spectral lines, we assume they arise from a minority impurity which does not contribute significantly to the continuum Bremsstrahlung spectrum. This is a good assumption in the case of a Hydrogen plasma, as in order to produce a 200eV x-ray, an electron must get much closer to the nucleus than the Bohr radius of the 13.6eV bound electron. From the perspective of a high-energy incident electron, even a neutral Hydrogen atom is an ion.

The most complete model of Bremsstrahlung emissions from partially ionized targets is partial-wave computation, but Avdonina and Pratt have produced a phenomenological model \cite{avdonina}. This model is mostly correct except for the XEDF emitted at energies near an electron shell energy. If a high-Z, partially ionized gas were the target, the correct procedure would be to use the Gaunt factor from Avdonina and Pratt, rather than the one given in Section \ref{sec:inversion}. We will explore high-Z target gas results in later publications. 

High-Z target gases of very small partial pressure may contribute to the Bremsstrahlung emissions a non-negligible amount. At higher energies than the k-shell energy, for example, the Hydrogen-like potential which causes the incident electron to emit Bremsstrahlung is enhanced by Z, the atomic number of the target nucleus. This enhances Bremsstrahlung emission by Z$^2$. A plasma which is $99\%$ hydrogen gas and $1\%$ water by partial pressure has $32\%$ of its Bremsstrahlung emissions above 543eV come from electrons incident on Oxygen nuclei. 

\section{Apparatus}
\label{sec:apparatus}

\begin{figure*}[tbp]
\centering\includegraphics[scale=0.45]{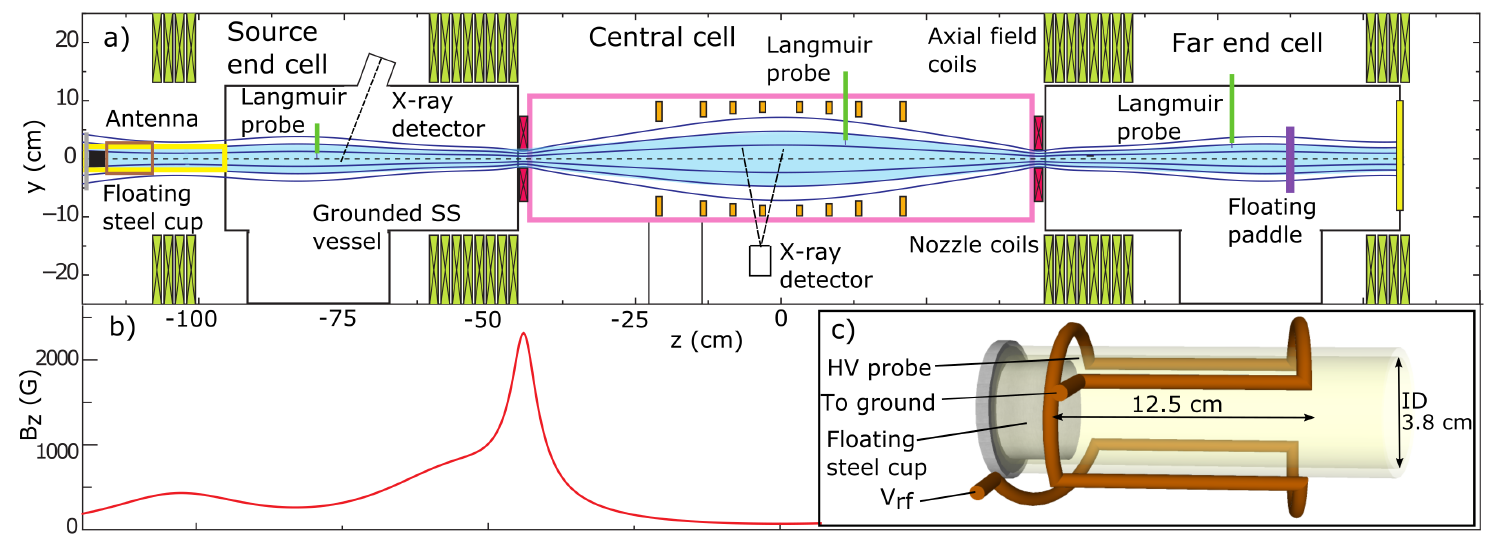}
\caption{Schematic of the PFRC-II device. The working gas is introduced in the Source End Cell (SEC), ionized by a double-saddle antenna, and flows through the double tandem mirror configuration into the Center Cell (CC). The magnetic field lines terminate on a floating Tantalum paddle in the Far End Cell (FEC).}
\label{fig:schematic}
\end{figure*}

The PFRC-II experiment is a magnetic confinement experiment at the Princeton Plasma Physics Laboratory.\cite{PFRC} It can be run either in Rotating Magnetic Field mode as an FRC, or in double tandem mirror mode as a high mirror ratio, low $\beta$ mirror. The experiments discussed here were performed with in the double tandem mirror mode. It is depicted in Figure \ref{fig:schematic}.

Gas is introduced into the Source End Cell (SEC) by an adjustable needle valve. The SEC contains a 30-cm-long, 3.8-cm inner-diameter Pyrex pipe with a 12-cm-long double-saddle antenna operated at 27 MHz and capable of exciting a helicon mode. In the experiments discussed here, the antenna is driven in a non-helicon mode, transmitting 200W-500W to the plasma, steady-state or square-wave modulated at 1-10 kHz. An electrically floating stainless steel cup terminates the Pyrex pipe. It is from this antenna and vacuum vessel arrangement that the superthermal electrons are produced, as discussed in a previous publication. \cite{jandovitz}

The SEC is connected to the Center Cell (CC) by a 2-cm inner diameter, 3-cm-long nozzle bore which is also the magnetic maximum. This magnetic maximum is operated typically around 2 kG. The CC is an 85-cm-long Lexan vacuum vessel with 23-cm inner diameter. The magnetic field at the midpoint of the CC is typically 70 G. 

On the downstream side of the CC is the Far End Cell (FEC), which contains a 5-cm-diameter tantalum paddle which is electrically floating. All field lines which traverse the nozzle bore terminate on this paddle. 

Each chamber is continuously pumped by a turbomolecular pump. Pressure ratios are typically 40:20:1 SEC:CC:FEC. Pressures are monitored by Baratron capacitive manometers and Granville Phillips ion gauges. Hydrogen is the typical working gas; argon, helium, and neon are also commonly used. The base pressure is below $10^{-6}$ Torr.

Magnetic fields are provided by independent, water cooled sets of nozzle and main (Helmholtz) electromagnetic coils. 

Langmuir probes in the SEC, CC, and FEC can be used to measure the bulk electron density and temperature in each chamber. Two Amptek X-100 SiPIN and one Amptek X-123 FAST SDD x-ray detectors can be mounted in the SEC and in the midplane of the CC. During the experiments described here, the SDD was mounted with line-of-sight to the midpoint of the CC. 

Typical bulk parameters for the tandem mirror mode are $T_e=4\text{eV}, n_e=1\cdot10^{11} /\text{cm}^3$. Maxwellian fits to the x-ray spectra yield parameters of the super thermal population of $T_e=300\text{eV}, n_e=3\cdot10^{9} /\text{cm}^3$ in the CC. XEDFs measured simultaneously in the SEC and CC show that the CC XEDF extends to higher energy and has a hotter effective temperature. 

\section{Results}
\label{sec:results}

\begin{figure}[tbp]
\centering\includegraphics[scale=0.45]{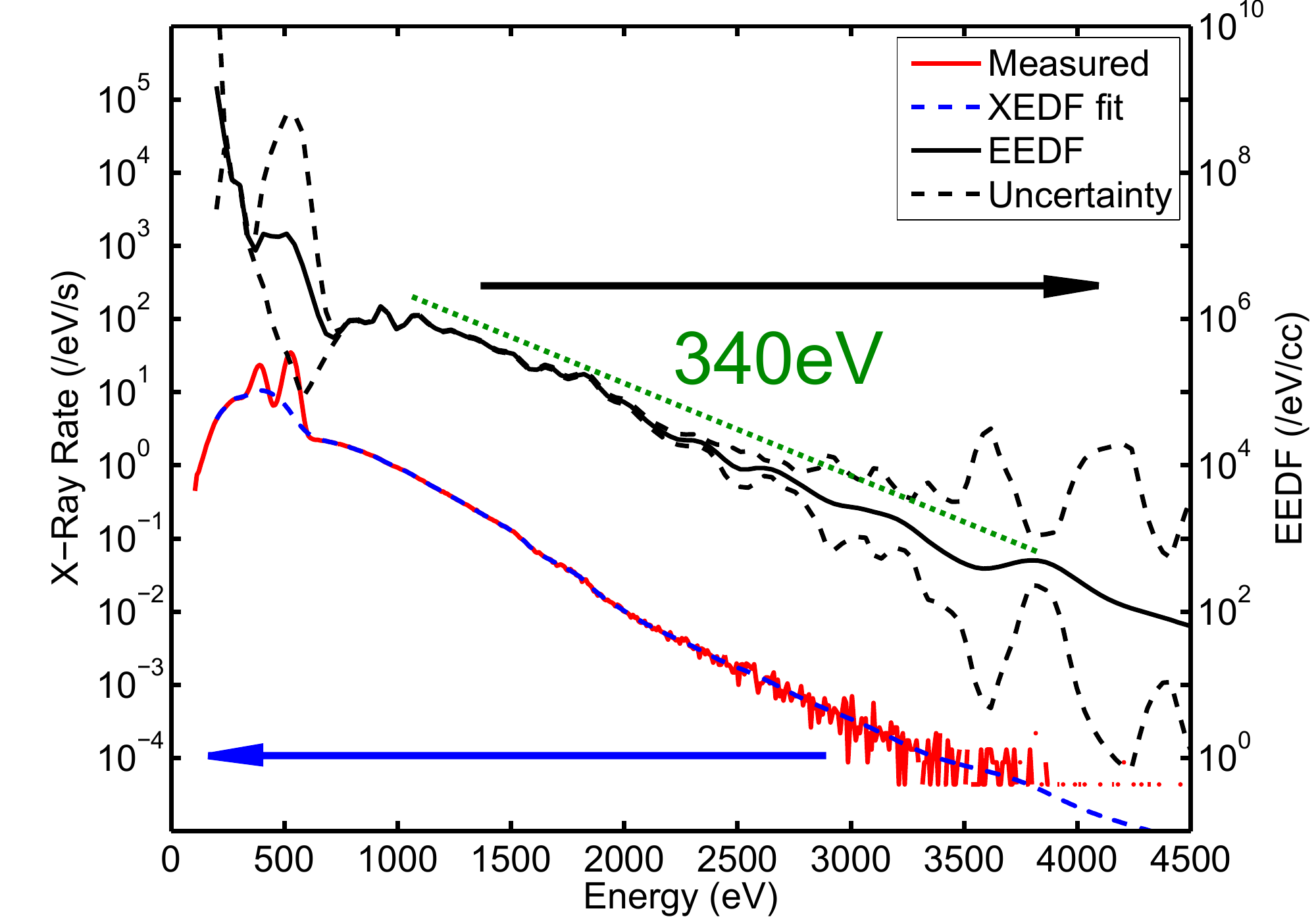}
\caption{A typical XEDF and EEDF from the CC in tandem mirror mode. The blue line is the measured XEDF, directly taken from the SDD. The black line and associated dotted lines are the derived EEDF and uncertainty, corresponding to $1\sigma$. The red line is this EEDF re-transformed into the expected measured XEDF for reasons of comparison. The EEDF is mostly exponential with an e-folding energy of 340 eV. The large uncertainty of the region around 500eV is caused by spectral lines obscuring Bremsstrahlung spectrum, excised as in Figure \ref{fig:wrong-spectral-lines}b.}
\label{fig:lowFECpress}
\end{figure}

One raw XDEF spectrum and its inverted EEDF are shown in Figure \ref{fig:lowFECpress}, obtained with 350 W of forward RF power being deposited into the antenna and hydrogen plasma. The gas pressure in the CC was 0.43 mTorr and $1.4\cdot 10^{-5}$ Torr in the FEC. The magnetic field in the midplane of the CC was 70 G and 2.2 kG at the nozzle. The floating potential of the tantalum paddle in the FEC was -1.2 kV.

The raw spectrum shows N and O K-$\alpha$ x-rays, necessitating the removal of that portion of the spectrum as in Section \ref{sec:spectral}. Figure \ref{fig:lowFECpress} shows a mostly exponential EEDF. This population of electrons could well be thermal with an effective temperature of 340 eV. If this Maxwellian distribution continued down to 0 eV, this population would have a density of $6.7\cdot 10^9/\text{cm}^3$. The low-energy portion of the spectrum is made uncertain by the presence of the N, O lines, but it appears that this Maxwellian behavior does not extend down to 500eV in EEDF, instead becoming ``colder'' and steeper at this energy.

By increasing the pressure in the FEC to $4.9\cdot10^{-5}$Torr, a factor of 3.5 larger than in Figure \ref{fig:lowFECpress}, the floating potential of the Tantalum paddle in the FEC becomes far less negative, -20 V, and the paddle glows cherry red hot. 

The increase in potential and power flow to the Tantalum paddle is indicative of a fast-electron created cold plasma in the FEC. Before the pressure was increased, the fast electron population was sufficient to balance the bulk ion current to the paddle, keeping its floating potential at -1.2 kV and keeping the power in the bulk plasma from heating the paddle. After the pressure was increased, we suspect that the fast electrons produced a plasma whose larger density was sufficient to set the paddle floating voltage to the measured -20 V, allowing power in the bulk plasma to heat the paddle. Characterizing this fast-ion produced plasma more fully with Langmuir probes will be the result of a later publication.

\begin{figure}[tbp]
\centering\includegraphics[scale=0.45]{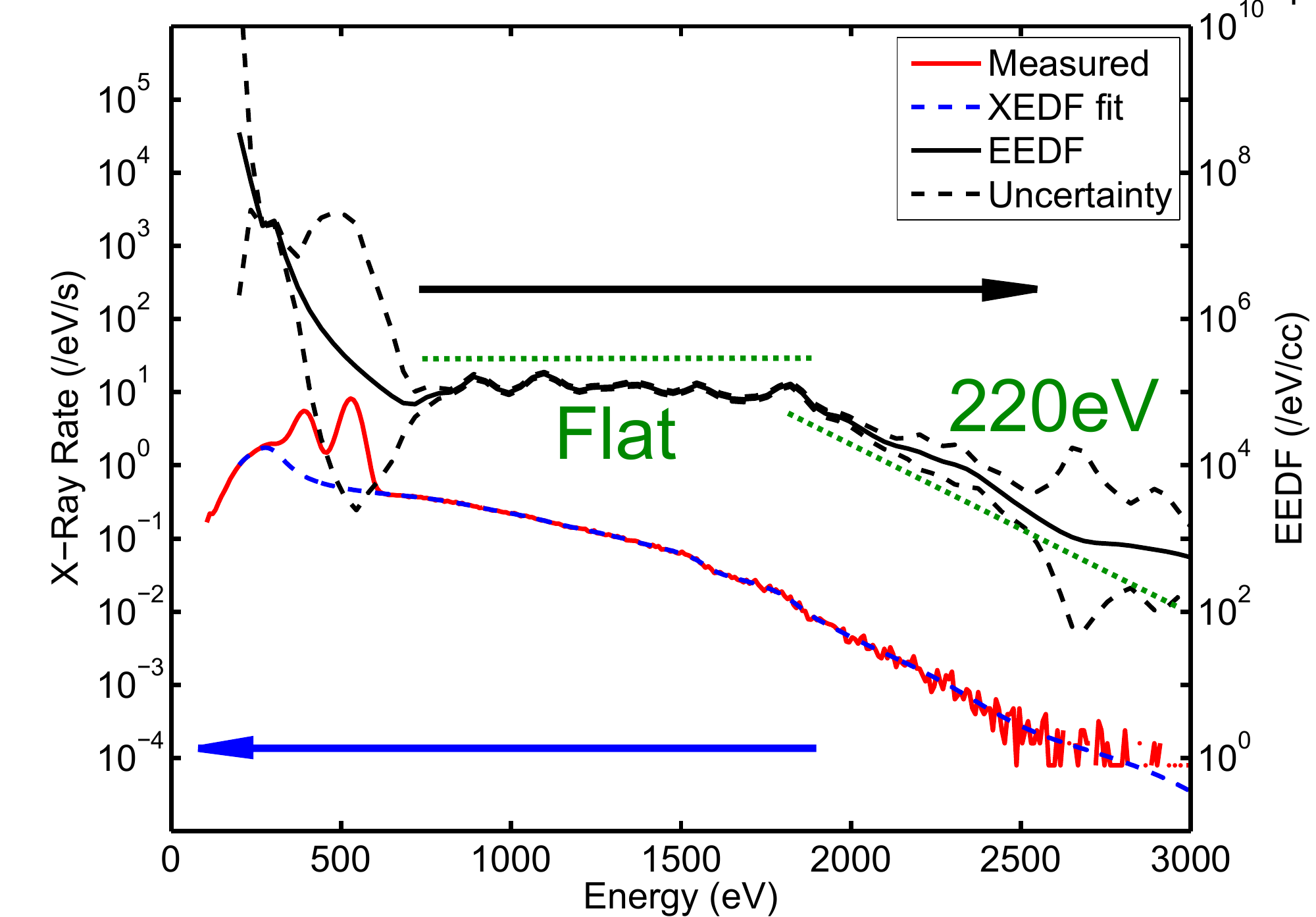}
\caption{XEDF and EEDF from the CC in tandem mirror mode with a high pressure in the far end cell. The blue line is the measured XEDF, directly taken from the SDD. The black line and associated dotted lines are the derived EEDF and uncertainty. The red line is this EEDF re-transformed into the expected measured XEDF for reasons of comparison. The EEDF is split into two domains: below 1800eV it appears flat. Above 1800eV it appears exponential with e-folding energy 220eV.}
\label{fig:highFECpress}
\end{figure}

The EEDF derived from this configuration is shown in Figure \ref{fig:highFECpress}. It displays a very different EEDF; at higher energies than 1800 eV, the EEDF has a much sharper fall than the low pressure condition, with an e-folding energy of 220 eV. This may be attributed to the effect of the Tantalum paddle no longer providing any electrostatic confinement of particles in the FEC. At energies lower than 1800 eV, the EEDF shallowed, appearing flat. Flattening of the EEDF can arise from collisional slowing-down of a beamlike distribution.\cite{hutchinson} More measurements are required to determine difinitively whether the flatness comes from acceleration of electrons at lower energy or slowing of electrons at higher energy.

\section{Conclusion}
\label{sec:conclusion}

We have developed an algorithm for the spectral inversion of Bremsstrahlung x-ray energy distribution functions (XEDFs) from plasma into electron energy distribution functions (EEDFs). It is better suited than the current state-of-the-art to low resolution, low count rate, low dynamic energy range conditions contaminated by spectral lines. Because the motivation for the algorithm is a log-likelihood optimization from $N$ Poisson variables, we call this inversion a Poisson-regularized inversion. 

This inversion has been used to measure the EEDF from a super-thermal population of electrons in the center cell (CC) of the PFRC-II device running as a low-power RF double tandem mirror. We find conditions of both thermal-like distributions and strongly non-thermal distributions.

Poisson-regularized spectral inversion will be used to probe the acceleration and dynamics of these super-thermal electrons in the PFRC-II device.

\subsection{Acknowledgment}

We would like to thank Bruce Berlinger for technical experimental assistance. We would also like to thank Manfred Bitter, Kenneth Hill, and Eugene Evans for helpful comments. This work was supported, in part, by DOE Contract No. DE-AC02-76-CHO-3073 and its Program in Plasma Science and Technology.

\end{document}